# Visualization of Interpersonal Communication using Indoor Positioning Technology with UWB Tags

Hayato Shinto[1], Yu Ohki,[2] Kenji Mizumoto[1]*, Kei Saito[1]*

[1] Graduate School of Advanced Integrated Studies in Human Survivability Kyoto University, Higashi-Ichijo-Kan, Yoshida-nakaadachicho 1, Sakyo-ku, Kyoto, 606-8306, Japan

[2] Faculty of Data Science, Rissho University, 1700 Magechi, Kumagaya-shi, Saitama, 360-0194, Japan

email: mizumoto.kenji.5a@kyoto-u.ac.jp, saito.kei.1y@kyoto-u.ac.jp

## ABSTRACT

In conjunction with a social gathering held on a university campus, the movement of attendees were tracked within the venue for approximately two hours using a UWB indoor positioning system, in order to visualize their interpersonal communication. Network and community analyses were performed on attendee interaction data, and the evolution of communities over time was further investigated through repeated community analysis at different time points. Furthermore, recognizing the influence of distance thresholds on defining contact, we discussed how varying these thresholds affected the resulting network structure and community analysis outcomes. This study confirmed that the temporal evolution of communities identified through community analysis broadly corresponded with the visually observed groupings of participants using the UWB indoor positioning system.

# 1. Introduction.

Interpersonal communication forms the basis of social interaction, and attempts have been made to visualize communication by accurately analyzing its frequency, scale, form, and changes. In particular, research related to visualization has been actively conducted in societies and organizations. A study conducted in Poland using a questionnaire survey targeting internal communication within a company reported that communication has a significant impact not only on a company's innovation efforts but also on the level of human resource utilization.(Cruz and Tavares, 2016) Another questionnaire survey in the same study highlighted the importance of visualizing regular communication in order to foster a common understanding and shared values.

The mainstream approaches of visualization and analysis for communication often relies on traditional approaches such as feedback surveys, focus group interviews, and qualitative analysis.(Meng and Pan, 2012) More recently, however, researchers have begun exploring the analysis of employee communication using quantitative methods by utilizing statistical data such as data from business chat applications.(Nonaka et al., 2023) The reason for this shift is that the communication analysis using questionnaire surveys is susceptible to recall bias. A recent study proposed using wearable IT devices and its automatic measurement has as an alternative method for the communication visualization. The advantages of this approach include the ability to simultaneously measure and analyze communication among a large number of individuals without requiring direct human observation, while also detect location coordinates, physical proximity, physical activity levels, and vocalizations.(Olguin et al., 2009) In that study, the data was further integrated with email communication data and survey responses to provide a multi-faceted analysis of organizational communication. Additionally, other studies have used location information obtained from IT devices such as Wi-Fi to analyze students' social behavior and academic performance.(Swain et al., 2023)

Other technologies other than Wi-Fi, such as Bluetooth low energy (BLE) beacons(Iseki and Hagiwara, 2018) are also applied for location tracking. However, the accuracy of location data obtained from these technologies can be limited due to positioning errors from the analysis, posing challenges for visualizing and analyzing interpersonal communication patterns.

Therefore, this study employed an ultra-wideband (UWB) indoor positioning system capable of high-precision positioning to visualize and analyze communication. Compared to commonly used indoor positioning technologies like Wi-Fi and BLE, which typically exhibits errors of several meters, UWB tags offer significantly improved accuracy, with positioning errors reduced to 10 to 50 centimeters (cm). Furthermore, UWB technology offers four key advantages. Firstly, it enables high-precision indoor positioning, as described above. Secondly, it facilitates anonymity and privacy protection. The UWB indoor positioning system used in this study achieves positioning by having each participant carry a UWB tag, which can be distributed randomly at the study's outset, ensuring anonymity. Moreover, the system collects only movement data without accessing sensitive personal information such as the content of activities or conversations, thereby protecting participant privacy. Thirdly, the UWB indoor positioning system enables three-dimensional indoor positioning. Fourthly, it allows for accurate indoor positioning even with a relatively large number of participants. By leveraging these features, This study was able to conduct large scale measurements and analysis of communication with guaranteed anonymity using UWB tags.

## 2. RELATED WORK

In recent years, UWB technology has emerged as a valuable tool for obtaining location information and has found applications in medical imaging, including techniques for visualizing organ structures within the human body(Klemm et al., 2009; Xu and Yang, 2008). Beyond these applications, UWB is also being implemented for physiological monitoring, such as heart rate measurement, and is now available in practical, commercial systems(Wu et al., 2019). In the field of location-based studies, numerous technical proposals and implementations demonstrate the feasibility of the use of UWB for highly accurate indoor positioning as mentioned before(Khoury and Kamat, 2009; Lee et al., 2007; Mazhar et al., 2017; Win and Scholtz, 1998; Yassin et al., 2017).

Much of the existing work on indoor positioning focuses on individual or small-group tracking, such as the studies monitoring elderly individuals in their homes. For instance, one study achieved sub-14 cm accuracy analysis of the individual monitoring in a home environment, suggesting potential application for monitoring the independent living style(Qian et al., 2024). While large-scale indoor positioning systems are utilized in industrial settings, such as factory wiring management, ongoing research continues to address challenges such as radio wave interference in densely populated environments. To our knowledge, there is a lack of research employing UWB for human positioning in large groups within real-world settings, and a comprehensive review of the academic literature reveals no prior studies of this nature.

Furthermore, although methodological discussions on the visualization of communication and community structures using indoor positioning data have emerged, empirical investigations utilizing UWB-obtained location data to identify these structures in real-world environments remain scarce(Iseki and Hagiwara, 2018; Ogata and Nakajima, 2020; Takahashi and Nakajima, 2020). Here we report the study addressing this gap by leveraging UWB tags to achieve high-precision indoor positioning within medium to large groups (approximately 30 or more participants) to visualize organizational communication and estimate community structures.

## 3. METHODOLOGY

### 3.1 DATA COLLECTION

Indoor positions of 26 research participants were recorded during an approximately two-hour standing social gathering. Participants were primarily trainees from corporate training programs at Kyoto University, who had prior acquaintance with one another. The gathering was conducted within a rectangular room. Individuals were informed about the function of the UWB tags and relevant privacy considerations, and written informed consent was obtained. During the social gathering, participants were instructed to engage in free-flowing interaction while consuming refreshments. No experimental manipulations or interventions were implemented.

In this study, we employed a UWB indoor positioning system provided by GIT Japan, Inc. The system comprises UWB anchors and UWB tags (Figure 1 and Figure 2, respectively). Eight UWB anchors were positioned within a rectangular experimental room measuring approximately 15 m x 7.8 m (Figure 3). Anchors were installed at the four corners and midpoints of the longer sides, at a height of approximately 2m above the floor. Each participant wore a UWB tag (dimensions: 4cm x 1.3cm x 7.9cm). Tags were either placed in the breast pocket of their clothing or secured to the chest using a pin. The system recorded location data at a frequency of 2 Hz (twice per second).

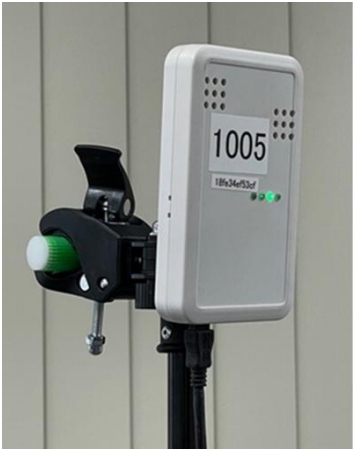
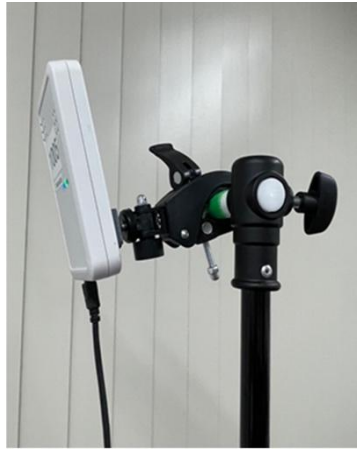
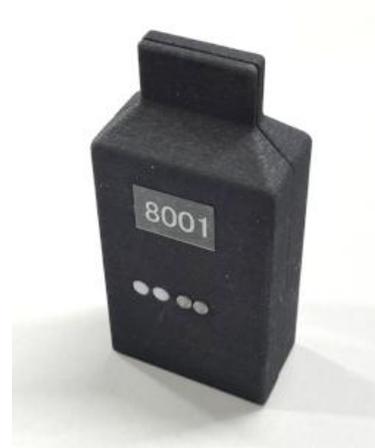

**Figure 1 UWB anchor**  **Figure 2 UWB tag**

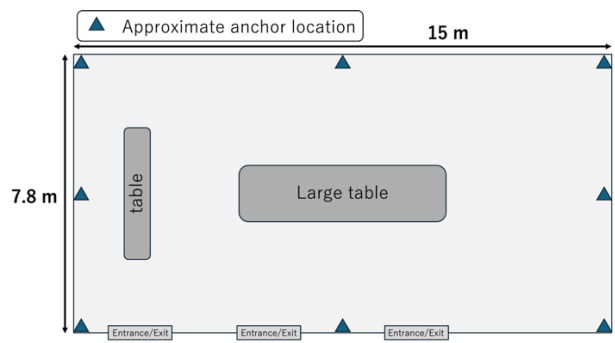

**Figure 3 Rough illustration of the room used for the experiment and approximate anchor location**

## 3.2 Preprocessing

### 3.2.1 Acquired Data & Initial Processing

The acquired data consisted of three-dimensional position coordinates for each of the 26 participants, recorded over approximately two hours under above condition. Participants' data from each UWB tag were then assigned anonymized IDs (P1-P40) for data analysis. Previous research has considered the impact of vertical (height) data on human communication patterns.(Ogata and Nakajima, 2020) However, given the standing, free-flowing nature of the experiment, we anticipated minimal influence of height variations on communication. Therefore, we reduced the dimensionality of the data by excluding the height component, utilizing only the two-dimensional (x, y) coordinates for subsequent analysis.

As initial processing, the data was sampled at a rate of two points per second and we used the average value of the these two points for further analysis.

### 3.2.2 DATA SMOOTHING AND IMPUTATION

While the system aimed for positioning updates twice per second, data gaps occurred when UWB tags were obscured from all anchors. Furthermore, the UWB indoor positioning system exhibited a measurement error of approximately 10-50cm, and initial data inspection revealed the presence of noise and missing values. As there is data gaps and errors from the data collection, previous research has applied and demonstrated the machine learning techniques, specifically Kalman filtering and Long Short-Term Memory (LSTM) models, to improve the accuracy of UWB-based positioning data.(Tian et al., 2024) In this study, we also employed both Kalman filtering and LSTM models to the obtained absolute coordinate data. The LSTM model utilized data from every preceding 120 seconds for every smoothing and imputation. The first 119 seconds of the whole experimental data were removed from the dataset prior to analysis as they were excluded from the LSTM-based smoothing and imputation process.

### 3.2.3 HANDLING PARTICIPANT ABSENCE AND ENTRY AND DISTANCE CALCULATION

During the experimental period, some participants briefly left the room (e.g., for restroom breaks) or joined the experiment from mid-session. Applying Kalman filtering and LSTM models without accounting for these events could introduce inaccuracies. To address this, we compared the raw data before and after smoothing and imputation. Any continuous data gaps exceeding 60 seconds in the raw data were identified as indicative of participant absence. The corresponding data segments for those participants were then treated as missing values in the smooth and imputed dataset.

We then calculated the Euclidean distance between each pair of participants based on their two-dimensional coordinates. This allowed us to quantify the proximity between each pair of the 26 participants within the room, forming the basis for our subsequent analysis.

## 3.4 DATA ANALYSIS

### 3.4.1 NETWORK CONSTRUCTION AND COMMUNITY ANALYSIS THROUGHOUT THE EXPERIMENT

Following data processing, we assessed the contact of the two participants based on the distance between pairs. Previous studies, and others relying on location data from technologies other than UWB, often use the technical limitations of the device as the distance threshold for analysis(Iseki and Hagiwara, 2018; Khoury and Kamat, 2009). However, the UWB indoor positioning system used in this study boasts positioning errors of approximately 10cm, allowing for highly accurate positioning that captures nearly all instances of human interaction. Therefore, directly applying the device's technical limitations as a distance threshold for analysis is not practical. Instead, we varied the threshold within the range where face-to-face communication is likely to occur, and then compared the resulting analytical outcomes. We established four thresholds – 0.75 m, 1.0 m, 1.25 m, and 1.5 m – and constructed networks for each, subsequently extracting and analyzing their respective network characteristics.

For example , with a threshold of 0.75 m, if the distance between a pair of participants was 0.75 m or less for a given second, a contact state of '1' was recorded; distances exceeding this threshold were recorded as '0'. The total number of '1's recorded for a given pair over the duration of the experiment represents the number of times that pair was within the threshold (total contact time). Based on this contact detection, we constructed a network representing contact relationships based on distance. The Edge-weighted Spring Embedded Layout was applied using each participants as nodes. The contact relationships of each threshold observed throughout the experimental period were converted into an edge list, which weight of each edges was defined total contact time between each participants. To clarify the structure of communication between the participants, we excluded from the edge list which contacts with a total contact time of less than 60 seconds as incidental(Ohki et al., 2023).

Next, we applied community analysis using Infomap(Edler et al., 2025) to each constructed networks to detect groups that communicate closely. The resulting network structure was visualized using the Edge-weighted Spring Embedded Layout in Cytoscape(Shannon et al., 2003) for each threshold. Pairs with zero contact time were excluded from the community analysis.

### 3.4.2 DYNAMIC NETWORK ANALYSIS

The static network analysis described above focused on total contact time between participants throughout the entire experiment. However, this approach does not capture how the interactions between participants and the formation of communication changed over time, as the static network is constructed from total contact time and does not distinguish between continuous and intermittent interactions. Consequently, it is unable to differentiate between frequent, brief exchanges (e.g., casual greetings) and sustained periods of close proximity contact.

To address this limitation, we constructed dynamic networks by dividing the experimental period into five-minute intervals. For each interval, an edge list and corresponding network were constructed, and community structure was identified using the same methodology as before. This allowed us to observe changes in interaction patterns over time.

The dynamic network analysis allowed us to visualize the frequency of contact within each interval, providing a more nuanced representation of interaction dynamics. Furthermore, we visualized the evolution of these communities over the entire experimental period using Alluvial diagrams getting by Plotly(Kruchten et al., 2025), tracking changes in community membership over time.

.

## 4. RESULT AND DISCUSSION

### 4.1 OVERVIEW OF ACQUIRED DATA

In this study, we tracked the indoor locations of 26 research participants using UWB tags during a two-hour standing social gathering. This resulted in 325 participant pairs. To characterize the distances between participants, we calculated the average location data for each participant over one-second intervals and then summarized the distances between all possible pairs as a histogram (Figure 4).

The resulting distribution ranged from the minimum measurement limit of approximately 1 cm to a maximum distance of approximately 14 m, with the overall tendency showing a peak around 3 m. This distribution is likely influenced by the dimensions (approximately 15 m x 7.8 m) and shape of the experimental room.

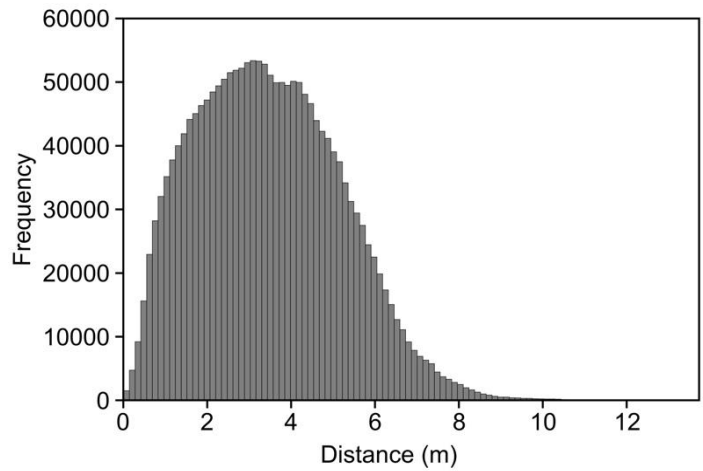

**Figure 4 Histogram of distances between all two points**

## 4.2 EXAMINING TRENDS FROM SELECTED PAIRS

Since the histogram above presents only the average distance between pairs, we randomly selected three pairs from the total of 325 pairs to analyze their distance data and investigate more detailed interaction trends. Figure 5 illustrates the changes in distance between these three pairs over the duration of the experiment. The notation such as p1_p22 in the figure indicates that it relates to a pair of P1 and P22. These data demonstrate that the distance between each pair fluctuated throughout the observation period. As an example, one of the selected pairs (P1_P22) experienced a change in distance from approximately 0.1 m to 8.9 m over the approximately two-hour period.

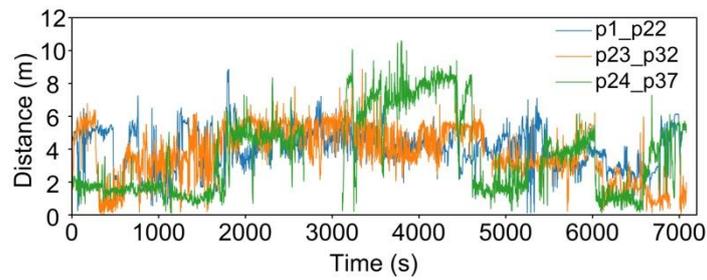

**Figure 5 Time course of distances for three randomly selected pairs**

To further characterize these interactions, we created histograms of all distance measurements for three pairs, with time on the x-axis and frequency on the y-axis. Comparing Figure 6 (A) to (C) to the overall distribution in Figure 4, we observe that several peaks are formed within each pair's histogram, unlike the broader distribution of all pairs. Notably, the location of the most prominent peak differed for each pair. This suggests that this distance of the peak may be indicative of their relationship and the position of these peaks may provide insights into the nature of the interaction between each pair.

The formation of multiple peaks in the histograms aligns with the observations in Figure 5, which demonstrate that the distance between participants fluctuated over time, indicating alternating periods of close proximity and separation. This suggests that communication patterns and community dynamics fluctuated over time, resulting in multiple shifts in the distance between participants.

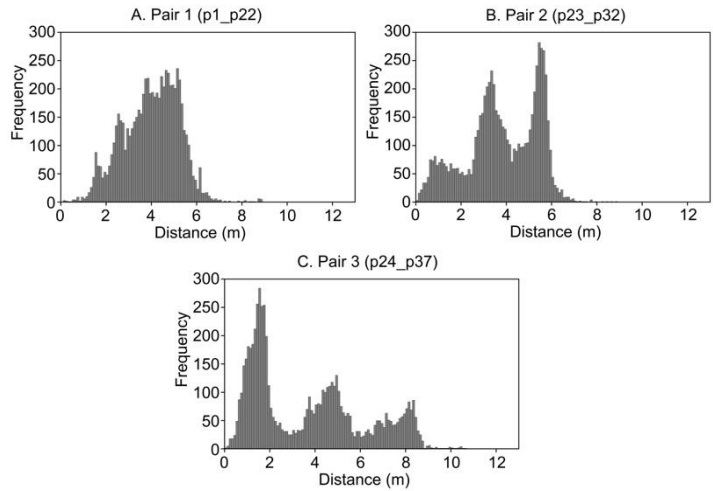

**Figure 6 Histogram of distances between randomly selected pairs over the experimental period, A(Histogram of distances between pairs of P1 and P22), B(Histogram of distances between pairs of P23 and P32), C(Histogram of distances between pairs of P24 and P37)**

## 4.3 Network Analysis of the Entire Experimental Period

First, we assessed the contact of the two participants based on the distance between pairs using distance thresholds of 0.75 m, 1.0 m, 1.25 m, 1.5 m and the assessment were made in one-second increments. Next, we constructed networks representing contact relationships between all participant pairs over the entire experimental period for each threshold. Figure 7 illustrates these networks, where each node represents a participant and the color of the edges connecting nodes indicates the frequency of contact (darker colors representing more frequent contact). The Edge-weighted Spring Embedded Layout was used, positioning nodes with more frequent contact closer together. Community information is not reflected in Figure 7 as the Infomap algorithm identified a single community across all thresholds. To compare network characteristics across thresholds, Figure 8 shows changes in assortativity (Figure 8–A), average clustering coefficient (Figure 8–B), average degree (Figure 8–C), and average shortest path length (Figure 8–D) for each threshold. All of these features varied depending on the threshold value. Notably, the assortativity showed significant changes between 1.0 m and 1.25 m, suggesting a substantial shift in network structure within this range.

These results indicate that, when analyzing the entire experimental period, participants generally interacted with a diverse range of individuals. However, there were few significant differences or biases in interactions between participants. This suggests that it is difficult to clearly identify communities or communication patterns using network and community analysis techniques for the analysis of the entire experimental period.

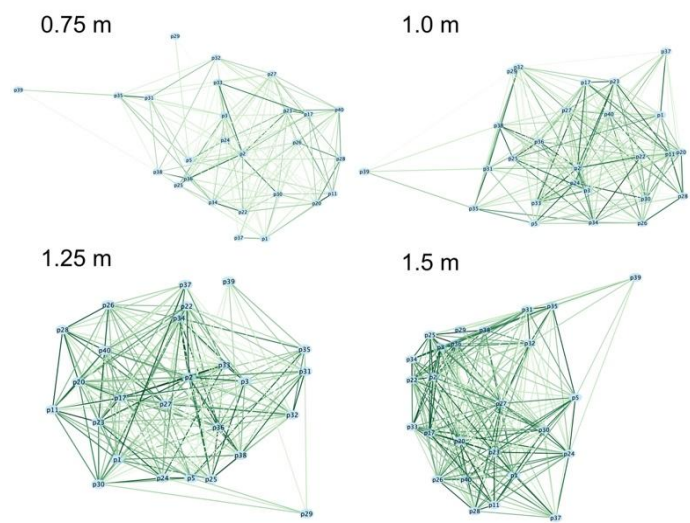

**Figure 7 Networks across the entire experimental period for each threshold**

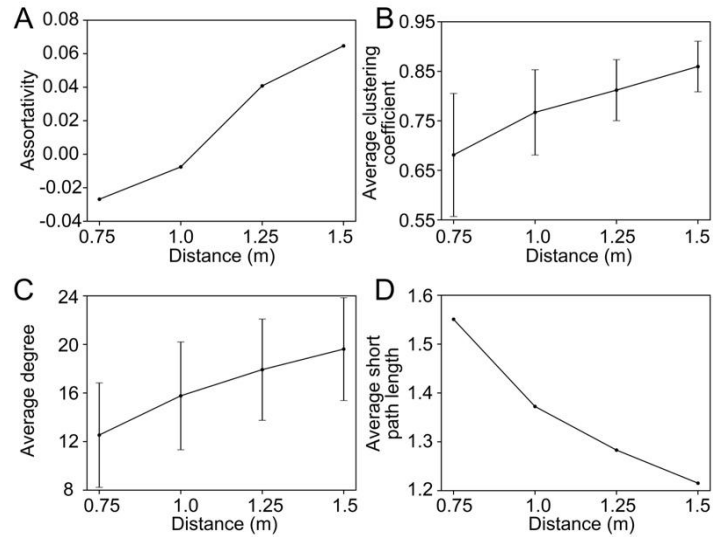

**Figure 8 Network characteristics for each threshold, A(assortativity), B(average clustering coefficient), C(average degree), D(average shortest path length)**

## 4.4 FIVE-MINUTE COMMUNITY ANALYSIS

We found difficulty conducting a comprehensive network analysis of the approximately two-hour location data as they showed obscure changes in spatial positioning and temporal contact patterns as described above. To address this, we divided the approximately two-hour dataset into 5-minute intervals, and performed community analysis on each interval using distance thresholds of 0.75 m, 1.0 m, 1.25 m, and 1.5 m. This generated a series of 24 networks, each representing a 5-minute period.

As an example, Figures 9-A and 9-C illustrate the networks generated for the first 5 minutes (0-299 seconds) and the subsequent 5 minutes (300-599 seconds) using a 1.0 m threshold. Edges in the figures are colored according to contact frequency (darker colors indicating more frequent contact), and nodes are colored based on their community assignment as determined by the Infomap algorithm.

Comparing these two networks reveals that some participants maintained consistent contact over the 5-minute interval, while others changed their contact patterns. The number of communities identified increased from four (Figure 9-A) to five (Figure 9-C).

To better visualize these changes in the communities, we used an Alluvial diagram (Figure 9-B) to track how individuals moved between communities over the two 5-minute intervals. The diagram shows community assignments for the first 5 minutes ([0s-1] to [0s-4]) and the subsequent 5 minutes ([300s-1] to [300s-5]).

Figure 9-B demonstrates that the initial four communities ([0s-1] to [0s-4]) evolved into five communities ([300s-1] to [300s-5]) over the 5-minute period. The diagram also visually highlights community splits and merges, as well as changes in community size. For example, the diagram shows that individuals who split from community [0s-1] joined with those who split from [0s-4] to form a new community, [300s-1]. Therefore, the Alluvial diagram effectively visualizes community dynamics, including splits, merges, and size changes.

To capture the temporal evolution of communities beyond the initial five minutes, we generated Alluvial diagram for all 24 networks, each representing a five-minute interval. Figure 10 illustrates this for a 1.0 m distance threshold, visualizing community structure changes from 0 to 1799 seconds. Similar to Figure 9-B, this diagram allows for a visual understanding of how communities evolved over time.

To further illustrate this dynamics, we tracked and monitored a specific participant (identified as P22) and also the community that he involved using a red line in Figure 10, showing the participant's movement between communities. This demonstrates the ability to track individual participant behavior within the evolving community structure based on UWB-derived location data.

We conducted this analysis across multiple distance thresholds, not just for 1.0 m but also for 0.75 m, 1.25 m, and 1.5 m, and presented the resulting Alluvial diagrams in Figure 11. While Figure 11 displays the data from 0 to 1799 seconds, the Alluvial diagrams for the remaining time period (1800 seconds onwards) are provided in the Supporting Information (Figure S1-S4).

The numbers displayed on each segment of the Alluvial diagrams represent the number of individuals belonging to a specific community during each five-minute interval. For example, the numbers "15, 5, 4" associated with the "0s~" segment (0-299 seconds) and a 1.5m threshold indicate that there were three communities at that time, with 15, 5, and 4 members respectively.

Comparing the Alluvial diagrams across different thresholds revealed that larger thresholds generally resulted in larger community sizes. However, the changes in community size were not substantial when comparing 0.75 m to 1.0 m, or 1.25 m to 1.5 m. In addition, the 0.75 m/1.0 m and 1.25 m/1.5 m pairs exhibited similar trends in community evolution.

These findings demonstrate that by adjusting the distance threshold, we can modify the perspective and resolution of our analysis when studying communities formed through face-to-face communication using UWB-based indoor positioning. This allows for a flexible approach to understanding the dynamics of these communities.

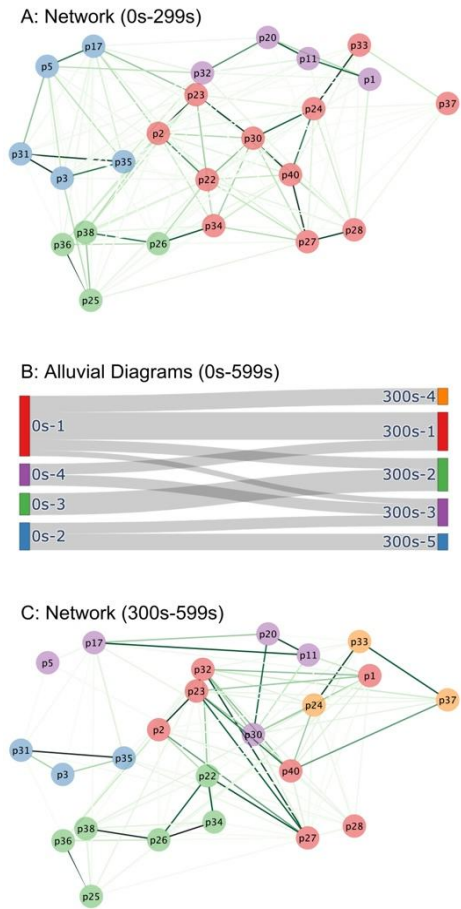

**Figure 9 Alluvial diagram illustrating the evolution of networks from 0-299 seconds and 300-599 seconds, and the community structure from 0-599 seconds, A(Network of 0s - 299s), B(Alluvial diagram of 0s – 599s), C(Network of 300s – 599s)**

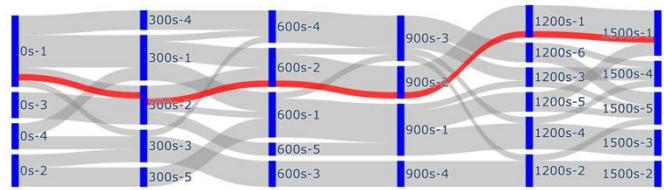

**Figure 10 Evolution of Infomap communities from 0-1800 seconds, showing the movement of a specific individual (P22) between communities**

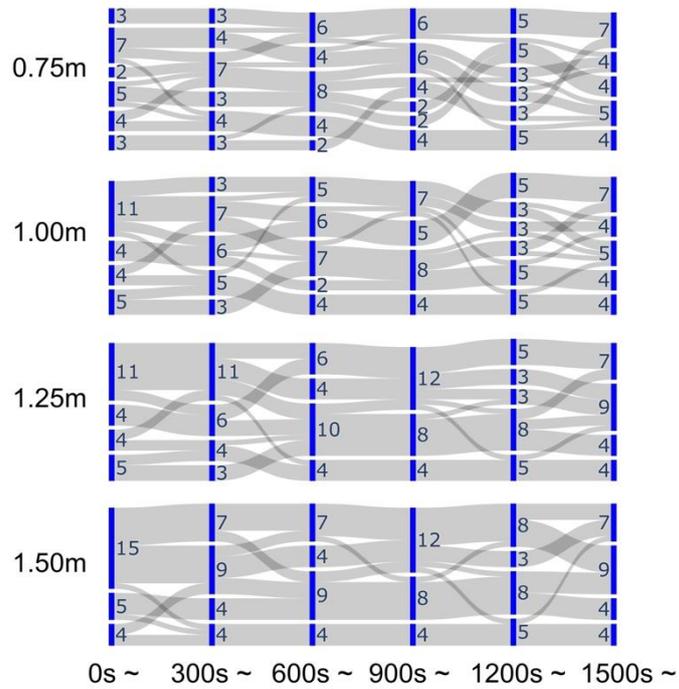

**Figure 11 Effect of threshold on the evolution of Infomap communities from 0-1800 seconds**

We have presented results based on network analysis using location data. Next, we focus on a key strength of the UWB indoor positioning system used in this study which is the ability to track the actual changes in each individual's physical location. We divided the experimental time into 5-minute intervals and compared the community information obtained through Infomap(Edler et al., 2025) with the corresponding contact states based on realistic positional coordinates. Figure 12 shows a plot of the positional coordinates of all participants from 0 to 299 seconds (Figure 12-A) and from 300 to 599 seconds (Figure 12–B and Figure 12-C). The plots in Figures 12-A and 12-C reflect the community information obtained at 0-299 seconds (same set as Figure 9-A), using color-coding. This allows us to visualize participant movement by comparing the plots of their actual positions. Comparing the two figures, we observed changes in positional arrangements that corresponded to the community changes shown in Figure 9-B.

Figure 12-C presents the same plot as Figure 12-B, but with the colors updated to reflect the community information via network analysis obtained from 300 to 599 seconds (from Figure 12). Comparing Figures 12-A and 12-C, we can see that participants moved during the 5-minute interval, and after moving, formed new community clusters with those nearby. Thus, by comparing Figures 12-A, 12-B, and 12-C, we can see that network analysis effectively tracks the changes in actual community structure as individuals move and the interaction between participants. For example, a participant who was part of the purple community between 0-299 seconds in Figure 12-A joined with a participant who was part of the red community in Figure 12-B, and together they formed a new red community between 300-599 seconds in Figure 12-C. This observation aligns with the explanation provided in Section 4.4 using Figure 9-B. The community transitions shown in Figure 9-B are based on contact frequency, however, we also observe similar merging and formation of communities in the actual movement of individuals, suggesting that discussing community transitions based on contact frequency is reasonably valid. Therefore, we confirmed that the community information obtained through network analysis in 5-minute intervals is well-correlated with the clusters of participants that can be observed from their spatial coordinates. In other words, this demonstrates that it is possible to analyze both the spatial coordinates and the temporal characteristics and changes of communities using network analysis based on 5-minute intervals.

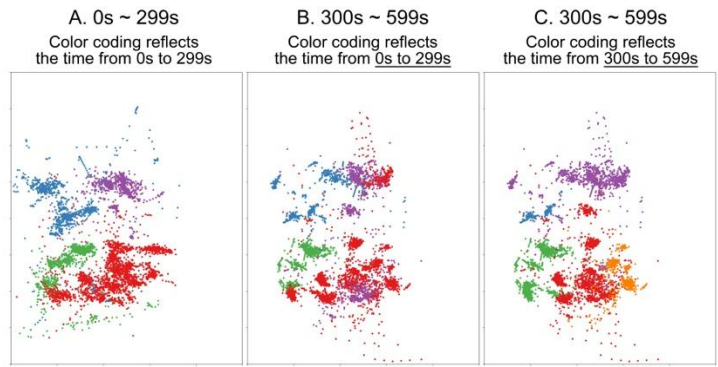

**Figure 12 Subject movement within each time interval and Infomap community structure, A(Plot of people's movements for 5 minutes from 0 to 299 seconds), B(Plot of people's movements for 5 minutes from 300 to 599 seconds), C(Plot of people's movements for 5 minutes from 300 to 599 seconds)**

## 4.5 Limitations of the Study

A limitation of this study is the lack of consideration for physical environmental factors when setting the threshold for network analysis. Specifically, the shape of the experimental room – a rectangle measuring approximately 15 m x 7.8 m – was a key physical environmental factor. The experiment utilized a standing reception format with a large table positioned centrally, which may have limited communication beyond the table's width. Consequently, participant distances appeared to be generally limited to approximately 4 meters, and this may be affected by the room's width. While a sufficiently large room was used, the 0.75 m-1.5 m threshold values were established without explicitly verifying their appropriateness. Furthermore, this study employed a single distance threshold to determine contact. In reality, communication can occur beyond 1.5 meters, and the nature and quality of communication may vary with distance. Therefore, it is important to note that this study only assessed communication and contact between individuals within our set values.

## 5. Conclusion

This study investigated a method for estimating community structure within a group using network analysis to visualize interpersonal communication in a limited indoor space, leveraging indoor positioning data obtained from UWB tags. Due to the high precision and high frequency (one data point per second) of the UWB data, we were able to analyze the changes in participant movement, transitions between communities, and shifts in community structure over time using the data segmented into 5 minutes intervals in this study. The results demonstrate that even short-term location data can be used to estimate and track community structure within a group.

Furthermore, while previous research has typically analyzed indoor positioning data based on the measurement limitations of the positioning system, this study leveraged the UWB tag's nominal minimum error of approximately 10 cm to explore various distance thresholds for contact detection, independent of system limitations. The results showed that changes in the distance threshold affected the size of the identified communities, indicating that the choice of threshold influences the results of community analysis.

## FUNDINGS

KM acknowledges support from the Japan Society for the Promotion of Science (JSPS) KAKENHI [Grant 20H03940 and 20KK0367]. HS acknowledges support from the JST SPRING [Grant JPMJSP2110] and the establishment of university fellowships towards the creation of science technology innovation [JPMJFS2123].

## DECLARATION OF COMPETING INTEREST

GIT Japan, Inc. provided an academic discount rate for UWB indoor positioning system rental.

# DECLARATION OF GENERATION AI IN SCIENCE WRITING

In this study, we received support from AI tools such as Gemini, ChatGPT, and GitHub Copilot including assistance with code development, data processing, proposing analysis content, and refining data processing and analysis methods. We utilized Gemma 3 and Gemini to support the translation of Japanese text into English. After using these tools, the authors reviewed and edited the content as needed and take full responsibility for the content of the published article.

# ACKNOWLEDGEMENT

We acknowledge GIT Japan, Inc. for providing the UWB indoor positioning system at an academic discount rate. We also gratefully acknowledge the participation of all individuals.

# REFERENCES

Cruz, A. C., and L. D. Tavares, 2016, Implementation of organizational climate survey for performance improvement and competitiveness of an information technology company: Sistemas & Gestao, v. 11, p. 290-298.

Edler, D., A. Holmgren, and M. Rosvall, 2025, The MapEquation software package.

Iseki, A., and T. Hagiwara, 2018, Analysis and Improvement for Communication between R&D teams using Indoor Positioning System, 2018 Spring National Research Presentation Conference of The Japan Society for Management Information, p. 142-145.

Khoury, H., and V. Kamat, 2009, Evaluation of position tracking technologies for user localization in indoor construction environments: Automation in Construction, v. 18, p. 444-457.

Klemm, M., I. Craddock, J. Leendertz, A. Preece, and R. Benjamin, 2009, Radar-Based Breast Cancer Detection Using a Hemispherical Antenna Array-Experimental Results: IEEE Transactions on Antennas and Propagation, v. 57, p. 1692-1704.

Kruchten, N., A. Seier, and C. Parmer, 2025, An interactive, open-source, and browser-based graphing library for Python.

Lee, J.-S., Y.-W. Su, and C.-C. Shen, 2007, A comparative study of wireless protocols: Bluetooth, UWB, ZigBee, and Wi-Fi: IECON 2007 - 33rd Annual Conference of the IEEE Industrial Electronics Society, p. 46-51.

Mazhar, F., M. G. Khan, and B. Sällberg, 2017, Precise Indoor Positioning Using UWB: A Review of Methods, Algorithms and Implementations: Wireless Personal Communications, v. 97, p. 4467-4491.

Meng, J., and P.-L. Pan, 2012, Using a balanced set of measures to focus on long-term competency in internal communication: Public Relations Review, v. 38, p. 484-490.

Nonaka, K., H. Yamashita, T. Miura, and M. Goto, 2023, Graph Embedding for Analysing Business Communication between Employees: IPSJ Journal, v. 64, p. 758-768.

Ogata, H., and T. Nakajima, 2020, Proposal of communication judgment analysis system for office using UWB communication, The 82nd National Convention of Information Processing Society of Japan, Kanazawa Institute of Technology Ohgigaoka Campus, Information Processing Society of Japan, p. 19~20.

Ohki, Y., H. Tanaka, and Y. Ikeda, 2023, Community Structure and Its Stability on a Face-to-Face Interaction Network in Kyoto City: Journal of the Physical Society of Japan, v. 92.

Olguin, D. O., B. N. Waber, T. Kim, A. Mohan, K. Ara, and A. Pentland, 2009, Sensible Organizations: Technology and Methodology for Automatically Measuring Organizational Behavior: Ieee Transactions on Systems Man and Cybernetics Part B-Cybernetics, v. 39, p. 43-55.

Qian, L., A. Chan, J. Cai, J. Lewicke, G. Gregson, M. Lipsett, and A. R. Rincón, 2024, Evaluation of the accuracy of a UWB tracker for in-home positioning for older adults: Medical Engineering & Phsics, v. 126.

Shannon, P., A. Markiel, O. Ozier, N. S. Baliga, J. T. Wang, D. Ramage, N. Amin, B. Schwikowski, and T. Ideker, 2003, Cytoscape: A software environment for integrated models of biomolecular interaction networks: Genome Research, v. 13, p. 2498-2504.

Swain, V. D., H. Kwon, S. Sargolzaei, B. Saket, M. B. Morshed, K. Tran, D. Patel, Y. Tian, J. Philipose, Y. Cui, T. Plötz, M. D. Choudhury, and G. D. Abowd, 2023, Leveraging WiFi network logs to infer student collocation and its relationship with academic performance: Epj Data Science, v. 12, p. 25.

Takahashi, Y., and T. Nakajima, 2020, Proposal and evaluation of identification function of office communication using UWB communication, The 82nd National Convention of Information Processing Society of Japan, Kanazawa Institute of Technology Ohgigaoka Campus, Information Processing Society of Japan, p. 21-22.

Tian, Y., Z. Lian, P. Wang, M. Wang, Z. Yue, and H. Chai, 2024, Application of a long short-term memory neural network algorithm fused with Kalman filter in UWB indoor positioning: Scientific Reports, v. 14.

Win, M. Z., and R. A. Scholtz, 1998, Impulse Radio: How It Works: IEEE Communications Letters, v. 2, p. 36-38.

Wu, S., T. Sakamoto, K. Oishi, T. Sato, K. Inoue, T. Fukuda, K. Mizutani, and H. Sakai, 2019, Person-Specific Heart Rate Estimation With Ultra-Wideband Radar Using Convolutional Neural Networks: IEEE Access, v. 7, p. 168484-168494.

Xu, H., and L. Yang, 2008, Ultra-wideband technology: Yesterday, today, and tomorrow: 2008 IEEE Radio and Wireless Symposium, p. 715-718.

Yassin, A., Y. Nasser, M. Awad, A. Al-Dubai, R. Liu, C. Yuen, R. Raulefs, and E. Aboutanios, 2017, Recent Advances in Indoor Localization: A Survey on Theoretical Approaches and Applications: IEEE Communications Surveys & Tutorials, v. 19, p. 1327-1346.

Supporting Information

# Visualization of Interpersonal Communication using Indoor Positioning Technology with UWB Tags

Hayato Shinto[1], Yu Ohki,[2] Kenji Mizumoto[1]*, Kei Saito[1]*

[1] Graduate School of Advanced Integrated Studies in Human Survivability Kyoto University, Higashi-Ichijo-Kan, Yoshida-nakaadachicho 1, Sakyo-ku, Kyoto, 606-8306, Japan

[2] Faculty of Data Science, Rissho University, 1700 Magechi, Kumagaya-shi, Saitama, 360-0194, Japan

email: mizumoto.kenji.5a@kyoto-u.ac.jp, saito.kei.1y@kyoto-u.ac.jp

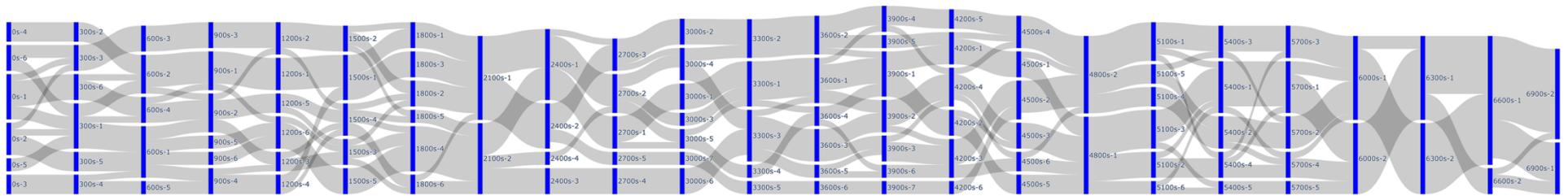

Figure S1 Alluvial diagram for the entire experiment when the threshold is 0.75m

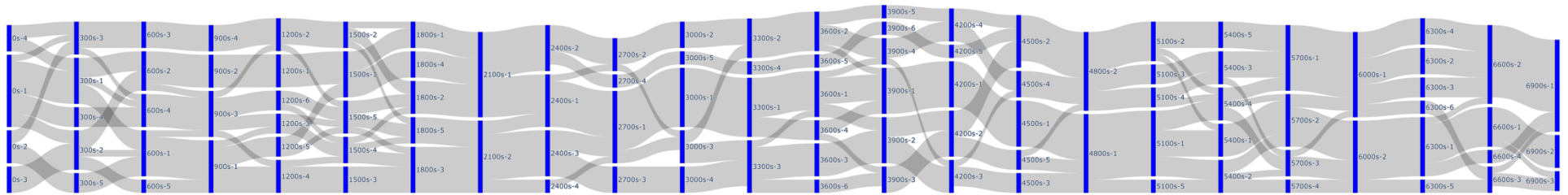

Figure S2 Alluvial diagram for the entire experiment when the threshold is 1.0m

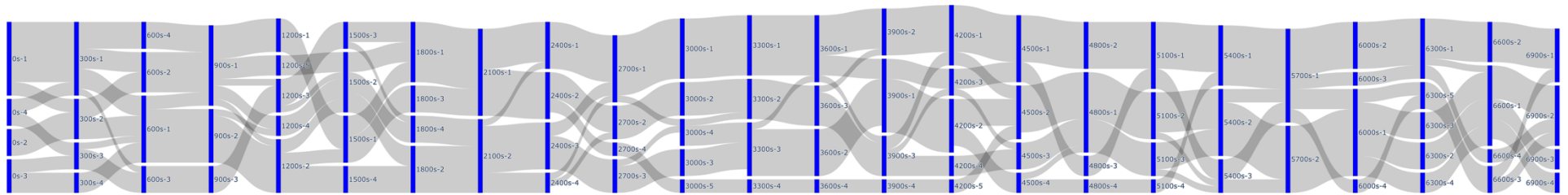

Figure S3 Alluvial diagram for the entire experiment when the threshold is 1.25m

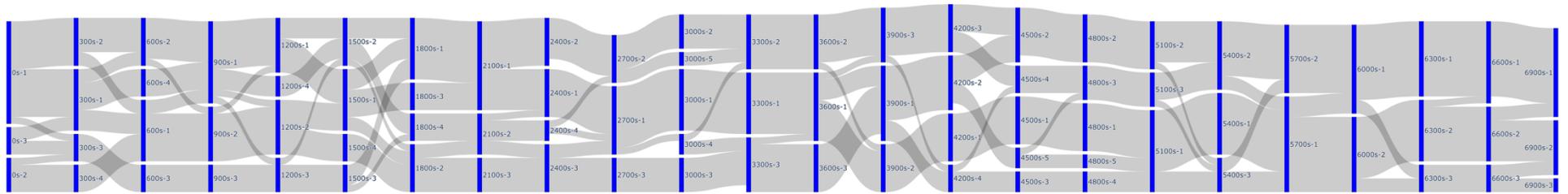

Figure S4 Alluvial diagram for the entire experiment when the threshold is 1.5m